\documentclass{article}
\usepackage{amssymb}
\usepackage{spconf,amsmath,graphicx,hyperref}
\usepackage{tikz}
\usetikzlibrary{arrows.meta,positioning,fit,calc}
\usepackage{setspace}
\usepackage{amsmath}
\title{Emotion and Acoustics Should Agree: Cross-Level Inconsistency Analysis for Audio Deepfake Detection}
\name{Jinhua Zhang \qquad Zhenqi Jia \qquad Rui Liu$^{\dagger}$ \thanks{$^{\dagger}$ denotes the corresponding author. This research was funded by the General Program(No.62476146) of the National Natural Science Foundation of China, the Young Elite Scientists Sponsorship Program by CAST(2024QNRC001), the Outstanding Youth Project of Inner Mongolia Natural Science Foundation(2025JQ011), the Key R\&D and Achievement Transformation Program of Inner Mongolia Autonomous Region(2025YFHH0014), and the Central Government Fund for Promoting Local Scientific and Technological Development(2025ZY0143).
} }
  
  \address{Inner Mongolia University\\
\texttt{zjh\_imu@163.com, jiazhenqi7@163.com, liurui\_imu@163.com}}
\begin{document}
%\ninept
%
\maketitle
\begin{abstract}
Audio Deepfake Detection (ADD) aims to detect spoof speech from bonafide speech. Most prior studies assume that stronger correlations within or across acoustic and emotional features imply authenticity, and thus focus on enhancing or measuring such correlations. 
However, existing methods often treat acoustic and emotional features in isolation or rely on correlation metrics, which overlook subtle desynchronization between them and smooth out abrupt discontinuities.
To address these issues, we propose EAI-ADD, which treats cross-level emotion–acoustic inconsistency as the primary detection signal. We first project emotional and acoustic representations into a comparable space. Then we progressively integrating frame- and utterance-level emotion features with acoustic features to capture cross-level emotion–acoustic inconsistencies across different temporal granularities. Experimental results on the ASVspoof 2019LA and 2021LA datasets demonstrate that the proposed EAI-ADD outperforms baselines, providing a more effective solution for audio anti-spoofing detection. The source code and demos are available at: https://github.com/AI-S2-Lab/EAI-ADD.
\end{abstract}
\section{Introduction}
\label{sec:intro}
% Deepfake technology leverages deep learning to synthesize non-existent audio-visual content—such as face swapping, voice cloning, face synthesis, and video generation—depicting people saying or doing things they never did. Its growing misuse for fake news and fraud makes effective detection imperative.
% Rapid advances in Text-to-Speech (TTS) \cite{liu2025retrieval,liu2025multi} and Voice Conversion (VC) \cite{yang2024streamvc,yao2025stablevc} have lowered the barrier to voice impersonation. 
Audio Deepfake Detection (ADD) aims to determine whether an audio sample is bonafide or spoof and is attracting growing interest \cite{liu2023betray,liu2024multi,liu2025hierarchical}. Recent work
leverages large audio models such as WavLM to obtain finegrained acoustic representations \cite{Combei2024WavLMME}. Some emotion-driven approaches primarily focus on utterance-level emotions and their correlations with other features \cite{conti2022deepfake,wu2024audio}.

% Existing methods largely fall into two lines: robust feature extraction and effective model design 

\begin{figure}[t]
  \includegraphics[width=\columnwidth]{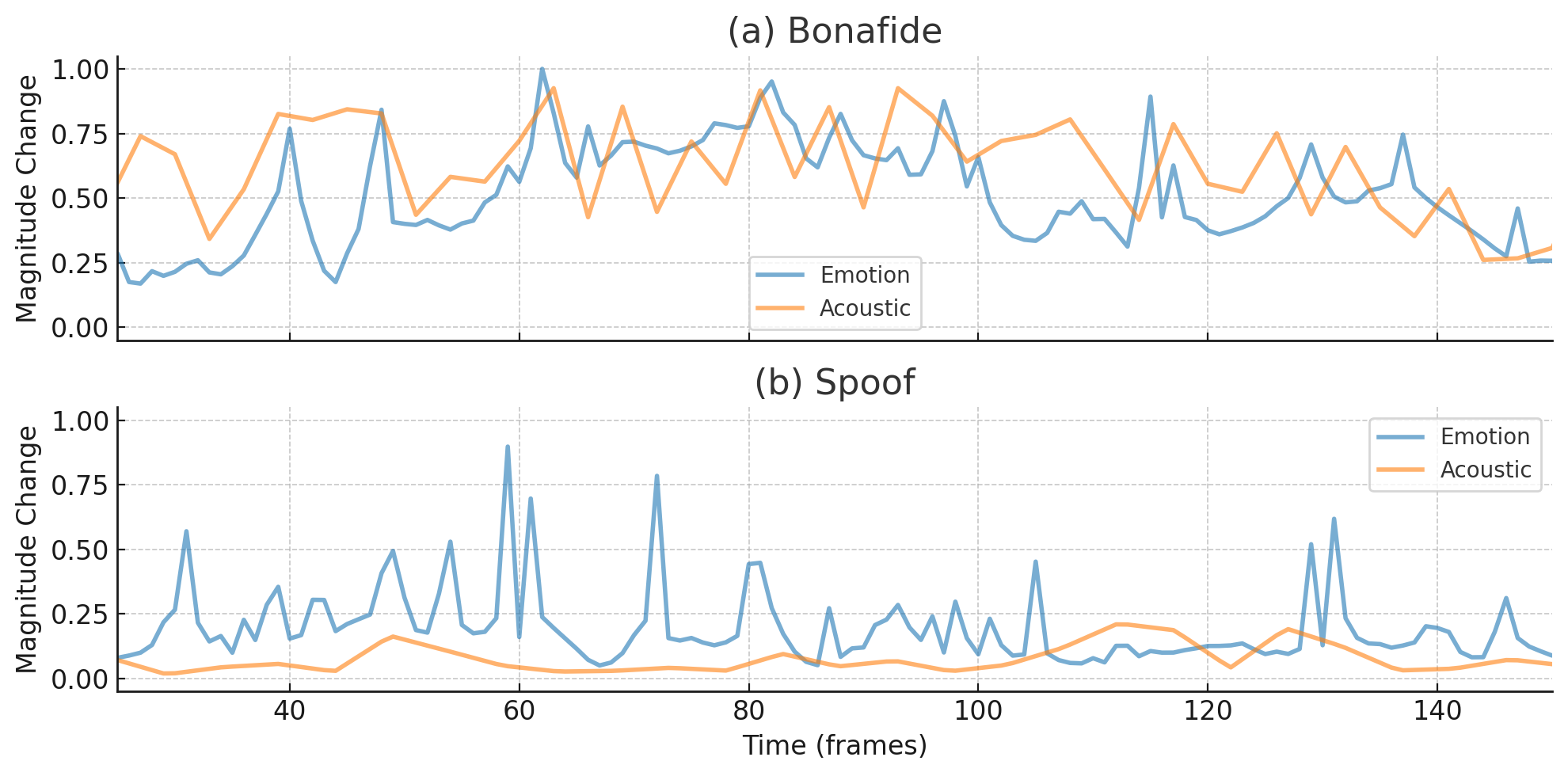}
  \vspace{-2.5em}
  \caption{
  Frame-level change magnitudes of emotion and acoustic features.
% (a) Bonafide speech exhibits smoother and more coherent co-variation patterns over time;
% (b) Spoof speech shows more irregular fluctuations with less consistent temporal correspondence.
  }
  \label{fig:1}
\end{figure}

% For the former, recent work leverages large audio models such as WavLM to obtain fine-grained acoustic representations~\cite{Combei2024WavLMME}.
% Some emotion-driven approaches primarily focus on utterance-level emotions and their correlations with other features. \cite{conti2022deepfake,wu2024audio}.
% For the second part, researchers adopt Gaussian mixture model, graph neural network, deep neural networks, recurrent neural networks etc. \cite{lei2020siamese,li2019multi,jung2022aasist} to build the ADD architecture. 

\begin{figure*}[t]
  \centering
  \includegraphics[width=\textwidth]{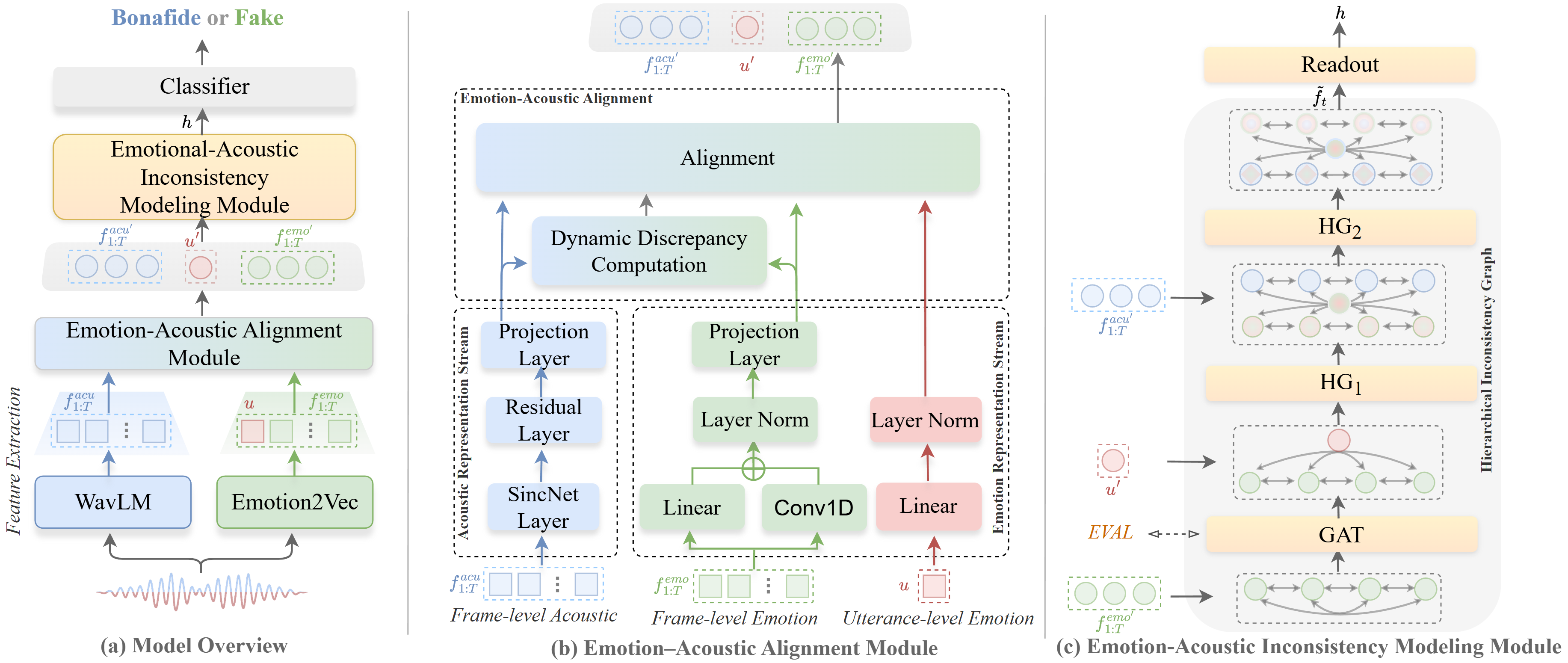}
  \vspace{-2.3em} 
  \caption{
  Our overall framework is illustrated in (a). 
  (b) illustrates the Emotion-Acoustic Alignment Module.
  (c) depicts Emotion-Acoustic Inconsistency Modeling Module.
  }
    \vspace{-1em} 
  \label{fig:2}
\end{figure*} 

Existing ADD works have largely analyzed patterns within individual acoustic or emotional features \cite{Combei2024WavLMME,zhang2025phoneme,wang2024ftdkd}, or measured how strongly the two are correlated \cite{wu2024audio}, using such correlations as cues to judge authenticity. However, in natural speech, emotional dynamics evolve smoothly over time and remain tightly aligned with the underlying acoustic structure \cite{scherer2003vocal,lin2023sequential,li2023music}. 
For example, a gradual increase in emotional arousal is usually accompanied by consistent variations in acoustic representations, whereas synthetic speech often shows prosody–emotion desynchronization, such as abrupt acoustic fluctuations without matching emotional changes \cite{chevi2024daisy,manku2025emergenttts}.
% As illustrated in Fig.~\ref{fig:1}, the top panel shows that bonafide speech maintains smooth and synchronized trajectories between emotional and acoustic features, whereas spoof speech exhibits clear desynchronization (e.g., sudden acoustic intensity rise without emotional change). 
% The bottom panel further quantifies this mismatch via the frame-level discrepancy $d_{\text{fra}}(t)$, which stays low and stable for bonafide speech but shows larger amplitudes and sharp spikes for spoof speech.
% These discontinuities break the expected temporal alignment between prosodic cues and emotion, making them reliable indicators of forgery. 
% Existing correlation-based analyses tend to smooth over or miss these subtle misalignments, treating them as statistical noise. We therefore shift the focus: rather than seeking correlation as a proxy for authenticity, we explicitly model mismatches between emotional dynamics and acoustic patterns as direct and discriminative cues, on the premise that such mismatches are rare in bonafide speech but prevalent in spoof audio.
As shown in Fig.~\ref{fig:1}, we use the L2 distance\footnote{L2 distance is chosen because it measures the magnitude rather than the direction of frame-level changes, making it well-suited to our task objective.} \cite{bishop2006pattern} to compute the frame-level change magnitudes of the emotion features and the acoustic features, and normalize them to \([0,1]\) for comparison on the same scale. In bonafide speech (Fig.~\ref{fig:1}(a)), the two curves co-vary over long stretches, indicating that acoustic fluctuations tend to co-vary with emotional dynamics over time.
In contrast, spoof speech (Fig.~\ref{fig:1}(b)) exhibits clear desynchronization: sharp bursts in the emotion curve are not accompanied by corresponding acoustic changes (and vice versa). 
Pearson correlations over 300 randomly selected samples also show more stable distributions for bonafide speech and greater inconsistency for spoof speech.
This difference in alignment patterns serves as the key cues of spoof speech. Existing correlation-based analyses tend to smooth over or miss these subtle misalignments, treating them as statistical noise. 
We therefore shift the focus: rather than seeking correlation as a proxy for authenticity, we explicitly model mismatches between emotional dynamics and acoustic patterns as direct and discriminative cues, on the premise that such mismatches are rare in bonafide speech but prevalent in spoof audio. 

To this end, this paper proposes a novel \textbf{E}motion–\textbf{A}coustic \textbf{I}nconsistency framework for ADD, termed \textbf{EAI-ADD}, that aims to uncover spoofing artifacts by modeling the inconsistency between emotional evolution and acoustic structural features. 
Specifically, we introduce an Emotion--Acoustic Alignment Module to project emotional and acoustic features into a unified representation space. 
On top of this alignment, the Emotion-Acoustic Inconsistency Modeling Module enforces reliable cross-level emotion–acoustic inconsistency analysis by integrating a Emotional Variation Amplification Loss to refine frame-level emotion trajectories and a Hierarchical Emotion–Acoustic Graph to link frame-/utterance-level emotional and frame-level acoustic nodes for detecting inconsistency patterns.
Experiments on the ASVspoof 2019 LA database demonstrate that our method consistently outperforms strong baselines. 
As far as we know, this is the first ADD study to explicitly model emotion--acoustic inconsistency as the key evidence for spoof detection.

\section{EAI-ADD Neural Architecture}
\label{sec:format}

We present EAI-ADD (Fig.~\ref{fig:2}), which consists of Feature Extraction Process, Emotion–Acoustic Alignment Module (EAAM), Emotion-Acoustic Inconsistency Modeling Module (EAIMM) and a Binary Classifier. 
The Feature Extraction process generates frame-/utterance-level emotional features and frame-level acoustic features. 
EAAM projects emotional and acoustic representations into a unified space. 
EAIMM enforces cross-level emotional inconsistency, capturing both short-term irregularities and cross-scale inconsistencies indicative of spoofing. The classifier produces the final bonafide/spoof decision.

\subsection{Feature Extraction}
We adopt WavLM to extract acoustic features for its stronger temporal-context modeling and fine-grained representations on short utterances compared with Wav2Vec2.0 \cite{Baevski2020wav2vec2A} and HuBERT \cite{Hsu2021HuBERTSS}, and we fine-tune it with AASIST \cite{jung2022aasist} for ADD by freezing lower layers and adapting upper layers to obtain task-relevant representations for accurate detection. 
We use frame-level acoustic features to capture subtle local anomalies, while utterance-level acoustic features are not used because global pooling may obscure such fine-grained discrepancies.  
Emotion2vec \cite{Ma2023emotion2vecSP} provides frame-/utterance-level emotional features. Its contrastive, context-aware pretraining effectively captures inter-frame emotional dynamics and yields superior emotion metrics such as lower MSE, making it well suited for our task. Frame-level emotion features reflect detailed dynamics, whereas utterance-level emotion features serve as contextual weights to aggregate and guide frame-level emotion differences, thereby providing a global reference for local emotional evolution.

\subsection{Emotion--Acoustic Alignment Module}
This module processes emotional and acoustic features through dedicated streams to mitigate redundancy while preserving key variation information, then projecting them into a unified representation.

\textbf{Acoustic Representation Stream (ARS).} 
This stream processes frame-level acoustic features $f^{acu}_{1:T}\in\mathbb{R}^{T \times d}$, focusing on compact representation without losing essential cues. 
Here, $f^{acu}_{1:T}$ denotes the sequence matrix formed by stacking the updated frame vectors $f^{acu}_{t}$.
The SincNet layer, designed as a parameterized band-pass 1D convolution, focuses on enhancing discriminative frequency components embedded in the acoustic representations.
The residual block then captures short-term dependencies and local fluctuations, safeguarding fine-grained acoustic dynamics.
Finally, a projection layer maps the features into a representation space compatible with emotion embeddings, providing stable inputs for alignment.

\textbf{Emotion Representation Stream (ERS).} 
This stream processes the frame-/utterance-level emotion features $f^{emo}_{1:T}\in\mathbb{R}^{T \times d}$ and $u\in\mathbb{R}^{1 \times d}$, preserving emotional information at different granularities.
Here, $f^{emo}_{1:T}$ denotes the sequence matrix formed by stacking the updated frame vectors $f^{emo}_{t}$.
In the frame-level pathway, the parallel operation of Linear and Conv1D aims to remove redundant information while preserving the ability to capture short-term dependencies and fine-grained emotional variations.
The summed output is normalized with LayerNorm and then projected to obtain stable frame-level emotional representations.
In the utterance-level pathway, the utterance-level emotion embedding is projected through a linear layer and normalized by LayerNorm, providing a stable representation of the global emotional trend.

\textbf{Emotion–Acoustic Alignment.} 
This part aims to align two kinds of features in a shared representation space:
\textbf{(1) Dynamic Discrepancy Computation:}
At the frame level, we first compute the discrepancy using first-order differences to capture the dynamic variations between consecutive frames over time. 
Concretely, 
$\Delta f_t = f^{\text{emo}}_{t} - f^{\text{emo}}_{t-1}$, 
$\Delta a_t = f^{\text{acu}}_{t} - f^{\text{acu}}_{t-1}$, 
and construct the frame-level change difference $d^{\text{fra}}=|\Delta f_t - \Delta a_t|$ as the dynamic discrepancy descriptor.
For the utterance level, average pooling is applied to the frame-level emotion features $u$ to match the dimension of the utterance-level embedding $u^{emo}$, the global discrepancy is computed as 
$d^{\text{utt}}=\big|u - u^{emo} \big|$.
\textbf{(2) Alignment:}
The descriptors $d^{\text{utt}}$ and $d^{\text{fra}}$ are fed into a fixed-sign dual-head softmax to produce two complementary weights:
$[\gamma_t^{\text{align}}, \gamma_t^{\text{mis}}] = \mathrm{softmax}([ -d^{\text{fra/utt}},\; +d^{\text{fra/utt}} ]).$
Here, $\gamma_t^{\text{align}}$ emphasizes alignment components, while $\gamma_t^{\text{mis}}$ explicitly encodes the degree of temporal inconsistency.
During feature updating, each stream selectively incorporates information from the other stream:
$f_t^{\text{emo}'/\text{acu}'} = \gamma_t^{\text{align}} \cdot f_t^{\text{emo/acu}} + \gamma_t^{\text{mis}} \cdot f_t^{\text{acu/emo}}$, $u{'} = \gamma_t^{\text{align}} \cdot u + \gamma_t^{\text{mis}} \cdot u^{\text{emo}}$.
The resulting features $f_{1:T}^{\text{emo}'}$, $u{'}$, and $f_{1:T}^{\text{acu}'}$ achieve alignment while retaining their respective fine-grained characteristics, enabling subsequent inconsistency detection.

\subsection{Emotion-Acoustic Inconsistency Modeling Module}
This module enforces reliable cross-level emotion–acoustic inconsistency analysis. 
EVAL employs contrastive learning to capture and amplify abnormal jumps in the temporal evolution of frame-level emotions, providing clearer discriminative emotion evolution cues.
HEIG employs the hierarchical heterogeneous graph modeling to progressively integrating frame- and utterance-level emotion features with acoustic structure features to capture cross-level emotion–acoustic inconsistencies across different temporal granularities.

\textbf{Emotional Variation Amplification Loss (EVAL).}
EVAL employs contrastive learning to capture and amplify abnormal jumps in the temporal evolution of frame-level emotions, providing clearer discriminative emotion evolution cues. 
We first compute the temporal difference $\Delta f_t^{emo}$ between frame $t$ and its immediate neighbor to characterize the local emotional transition at time $t$. 
To obtain a reference for this transition, we aggregate the temporal differences 
$\{\Delta f_j^{emo} \mid j \in \mathcal{N}(t)\}$ from the neighboring frames into a prototype $g_t$, 
where the sentence-level embedding $u{'}$ provides contextual weights. 
Here, $\mathcal{N}(t) = \{\, j \mid 0 \le j \le T-2,\ |j - t| \le k \,\}$ 
denotes the set of indices within a temporal window of radius $k$ around $t$, 
with $k$ being a predefined window size:
\begingroup
\setlength{\abovedisplayskip}{4pt}
\setlength{\belowdisplayskip}{4pt}
\begin{equation}
g_t = \frac{\sum_{j \in \mathcal{N}(t)} \alpha_{tj}\,\Delta f_j^{emo}}{\sum_{j \in \mathcal{N}(t)} \alpha_{tj}}, 
\quad 
\alpha_{tj} = \exp\!\left(\frac{u{'} \cdot f_j^{(1)}}{\tau}\right).
\end{equation}
\endgroup
Here, $g_t$ represents the expected local trend of emotional variation under the global guidance of $u{'}$, $f_j^{(1)}$ will introduced in HIG. 
We then apply a contrastive objective InfoNCE \cite{chen2020simple} to enforce $\Delta f_t$ to align with its prototype while being separated from negative samples $\Delta f^-$.
Specifically, negative samples $\Delta f^-$ are obtained from two sources within the same utterance: (i) temporal differences between far-apart frames to break short-term continuity; (ii) temporal differences from randomly shuffled frame orders to destroy natural sequential evolution. 

\textbf{Hierarchical Inconsistency Graph (HIG).}
HIG captures cross-level emotion–acoustic inconsistencies by progressively integrating frame- and utterance-level emotion representations with acoustic structure via a multi-stage heterogeneous graph.
We first construct a frame-level temporal graph, where each frame node aggregates information from its temporal neighbors via graph attention:
{\setlength{\abovedisplayskip}{3pt}
 \setlength{\belowdisplayskip}{3pt}
\begin{equation}
\{ f_t^{(1)} \}_{t=1}^T
= \mathrm{GAT}\!\big(f_t^{\text{emo}'}, \{f_{t-1}^{\text{emo}'}, f_{t+1}^{\text{emo}'}\}\big),\ t\!=\!1,\dots,T .
\end{equation}
}
Here, $f_t^{(1)}$ denotes the frame-level emotion feature after the first-stage GAT update, encoding local temporal context.
On top of this, EVAL amplifies abnormal temporal jumps, suppressing natural transitions and preserving salient inconsistency cues.
The enhanced frame embeddings are then integrated with the utterance-level emotion node $u'$ and frame-level acoustic nodes $w_t$ to form the node set.
$V = \{ f_t^{(1)} \}_{t=1}^T \cup u{'} \cup \{ f_t^{\text{acu}'} \}_{t=1}^T$,
where the edge set includes $E_{FF}$, $E_{FU}$, and $E_{FS}$. 
All cross-level connections are implicitly modeled by graph attention, without manually specified edge rules.
The update is performed in two heterogeneous stages: the first stage (\(\mathrm{HG}_1\)) detects local--global emotional inconsistencies, and the second stage (\(\mathrm{HG}_2\)) detects emotion--acoustic mismatches:
{\setlength{\abovedisplayskip}{3pt}
 \setlength{\belowdisplayskip}{3pt}
\begin{equation}
\tilde{f}_t = HG_2\!\left( HG_1\!\left( \{ f_t^{(1)} \}_{t=1}^T, u' \right), \{ f_t^{\text{acu}'} \}_{t=1}^T \right)
\end{equation}
The node-level outputs $\tilde{f}_t$ are aggregated by the pooling-based Readout, producing a compact graph-level representation $h$ that is subsequently fed into the binary classifier.
}

\begin{samepage}
\subsection{Overall Objective}
The final loss combines classification cross-entropy (CE) and EVAL using homoscedastic uncertainty weighting~\cite{kendall2018multi}. 
Let $s=\log\sigma^{2}$ be trainable: $ \mathcal{L} = \mathcal{L}_{\mathrm{CE}} + e^{-s}\mathcal{L}_{\mathrm{EVAL}} + s ,$
where $s$ regularizes the weighting to prevent vanishing weights.
\end{samepage}

\section{Experiments}
\subsection{Experimental Protocol}
% To ensure consistency with prior work, we evaluate our method on the ASVspoof 2019LA \cite{todisco2019asvspoof} and ASVspoof 2021LA \cite{yamagishi2021asvspoof} datasets. The 2019 set contains 6 known attack types for training and development, and 13 types (2 known, 11 unknown) for evaluation. The 2021 sets introduce unseen attacks and signal distortions under zero-shot conditions, with models trained on the 2019 set, thus offering a stricter test of generalization.
% Performance is assessed using Equal Error Rate (EER) \cite{wu2015asvspoof} and tandem Detection Cost Function (t-DCF) \cite{kinnunen2018t}.
% All audio samples are standardized to a fixed length of 64,600 samples by truncation or zero-padding, ensuring batch-level consistency while preserving the original signal onset. 
% The model contains 0.83M trainable parameters and requires 0.34 GFLOPs per utterance.
% It is trained for 60 epochs using the Adam optimizer with a learning rate of $10^{-5}$ and a weight decay of $10^{-4}$ on an NVIDIA A100 GPU.
% Hyperparameters are tuned on the validation set, and reported results are averaged over three random seeds. 
To ensure consistency with prior work, we evaluate our method on the ASVspoof 2019LA \cite{todisco2019asvspoof} and 2021LA \cite{yamagishi2021asvspoof} datasets.
% ASVspoof 2019LA contains 6 known attacks for training/development and 13 attacks (2 known, 11 unknown) for evaluation, while ASVspoof 2021LA introduces unseen attacks under a zero-shot setting, providing a stricter generalization test.
Performance is evaluated using EER \cite{wu2015asvspoof} and t-DCF \cite{kinnunen2018t}.
Audio samples are standardized to 64,600 samples.
The model has 0.83M parameters, requires 0.34 GFLOPs per utterance, and is trained for 60 epochs using Adam ($10^{-5}$ learning rate, $10^{-4}$ weight decay) on an NVIDIA A100 GPU, with results averaged over three random seeds.

\subsection{Main Results}
% The evaluation results in Table \ref{comparison} compare our proposed EAI-ADD with existing approaches on 2019LA evaluation set. Our method achieves a min t-DCF of 0.0110 and an EER of 0.34\%, representing the best reported results. Compared with the strongest baseline, EAI-ADD reduces t-DCF by 4.3\% and EER by 15\%, demonstrating its effectiveness and superiority. On the 2021LA dataset, EAI-ADD also achieves consistent improvements, indicating stronger generalization to unseen spoofing conditions.  
% These performance gains mainly stem from the design of EAI-ADD in modeling emotion–acoustic inconsistency. By aligning emotional and acoustic features in a unified representation space, the model can better capture subtle differences between the two. On this basis, EAI-ADD not only strengthens the detection of abnormal variations in frame-level emotion trajectories but also leverages cross-level graph relations to highlight mismatches between emotional evolution and acoustic structures. This dual constraint enables the model to maintain strong discriminative power even under unseen attacks and complex distortions, thereby achieving significant improvements over existing methods on both datasets.
Table~\ref{comparison} reports the performance of EAI-ADD against existing approaches on the 2019LA evaluation set.
EAI-ADD achieves the best results, with a min t-DCF of 0.0110 and an EER of 0.34\%, reducing t-DCF and EER by 4.3\% and 15\% compared with the strongest baseline.
Consistent improvements are also observed on 2021LA, indicating stronger generalization to unseen spoofing conditions.
These improvements stem from EAI-ADD’s modeling of emotion–acoustic inconsistency. By aligning emotional and acoustic features and leveraging cross-level graph relations, the model highlights abnormal emotion variations and emotion–acoustic mismatches, maintaining strong discriminative power under unseen attacks and complex distortions.

\subsection{Ablation Study}
Table~\ref{tab:ablation_all} presents ablation results on the ASVspoof 2019LA evaluation set.
\textbf{(1) Main Component Ablation.}
Removing any core module degrades performance. EAAM removal increases t-DCF/EER to 0.0118/0.42\%, EVAL removal yields 0.0117/0.41\%, and removing HIG causes the largest drop (0.0121/0.44\%), underscoring the role of multi-level consistency reasoning.
\textbf{(2) Alternative Mechanism Variants.}
Replacing ARS and ERS with linear projection leads to a clear performance drop (0.0125 / 0.46\%), indicating that naive dimensionality reduction fails to capture fine-grained or complementary emotion–acoustic patterns.
Similarly, replacing GAT with GCN further degrades performance (0.0133 / 0.52\%), which highlights the necessity of adaptive attention-based aggregation for effectively modeling heterogeneous relationships.
\textbf{(3) Hyperparameter Study ($k$ in EVAL).}
A temporal window size of $k=3$ achieves the best performance (0.0110 / 0.34\%). Both smaller ($k=1,2$) and larger ($k=4,5$) windows yield inferior results, indicating that a moderate temporal scope best balances short-term dynamics and noise suppression.
Overall, these ablation results confirm that each module is essential. Simplifying ARS/ERS to linear projection or replacing GAT with GCN consistently degrades performance, underscoring the importance of dedicated streams and attention-based message passing for modeling fine-grained and heterogeneous dependencies.
{
\setlength{\textfloatsep}{6pt}   % 正文 <-> 表格
\setlength{\floatsep}{6pt}       % 表格 <-> 表格

\begin{table}[t]
\centering
\caption{Comparison with other anti-spoofing systems on 2019LA and 2021LA datasets (pooled min t-DCF and EER). (– means unreported due to closed source baseline.)}
\label{comparison}
{\fontsize{7.5pt}{9pt}\selectfont
\renewcommand{\arraystretch}{0.9}
\setlength{\tabcolsep}{1.2pt}
\begin{tabular}{p{3cm}cccc}
\hline
\textbf{System} & \multicolumn{2}{c}{\textbf{2019 LA}} & \multicolumn{2}{c}{\textbf{2021 LA}} \\
& \textbf{t-DCF} ($\downarrow$) & \textbf{EER(\%)} ($\downarrow$)
& \textbf{t-DCF} ($\downarrow$) & \textbf{EER(\%)} ($\downarrow$) \\
\hline
SE-Rawformer \cite{liu2023leveraging} & 0.0344 & 1.05 & 0.3186 & 4.98 \\
AASIST \cite{jung2022aasist} & 0.0275 & 0.83 & 0.3398 & 5.59 \\
DFSincNet \cite{huang2023discriminative} & 0.0176 & 0.52 & 0.2731 & 3.38 \\
f0+Res2Net \cite{fan2024spatial} & 0.0159 & 0.47 & 0.2642 & 3.61 \\
WavLM+MFA \cite{guo2024audio} & 0.0126 & 0.42 & - & 5.08 \\
WavLM+RAD-MFA \cite{kang2024retrieval} & 0.0115 & 0.40 & - & 4.83 \\
\hline
\textbf{EAI-ADD (Ours)} & \textbf{0.0110} & \textbf{0.34} & \textbf{0.2533} & \textbf{3.29} \\
\hline
\end{tabular}
}
\end{table}

\vspace{-3mm}

\begin{table}[t]
\centering
\caption{Ablation study on the 2019LA dataset. 
The upper block removes key modules, the middle block replaces streams and graph variants, and the lower block studies the effect of temporal window size $k$ in EVAL.}
\label{tab:ablation_all}
{\fontsize{7.5pt}{9pt}\selectfont
\renewcommand{\arraystretch}{0.9}
\begin{tabular}{p{4cm}cc}
\hline
\textbf{Setting} & \textbf{t-DCF} $(\downarrow)$ & \textbf{EER(\%)} $(\downarrow)$ \\
\hline
\multicolumn{3}{l}{\textit{Main Component Ablation}} \\
\hline
w/o EAAM & 0.0118 & 0.42 \\
w/o EVAL & 0.0117 & 0.41 \\
w/o HIG  & 0.0121 & 0.44 \\
\hline
\multicolumn{3}{l}{\textit{Alternative Mechanism Variants}} \\
\hline
ARS\&ERS $\rightarrow$ Linear Projection & 0.0125 & 0.46 \\
GAT $\rightarrow$ GCN & 0.0133 & 0.52 \\
\hline
\multicolumn{3}{l}{\textit{Hyperparameter Study ($k$ in EVAL)}} \\
\hline
$k=1,2$ & 0.0120 / 0.0115 & 0.43 / 0.39 \\
$\boldsymbol{k=3}$ & \textbf{0.0110} & \textbf{0.34} \\
$k=4,5$ & 0.0118 / 0.0123 & 0.41 / 0.46 \\
\hline
\end{tabular}
}
\end{table}

} % ===== 局部设置结束 =====

\section{Conclusions}
This paper presented an ADD framework EAI-ADD that explicitly models emotion–acoustic inconsistency as the primary discriminative cue. The proposed EAAM aligns emotional and acoustic representations within a unified space, while the EAIMM conducts hierarchical inconsistency analysis to capture cross-level discrepancies. Experimental results on two datasets demonstrate that EAI-ADD achieves best performance. 
Notwithstanding these gains, the approach has yet to be validated under real-time or highly compressed conditions, which will be the focus of future investigations.

{\footnotesize
\bibliography{strings,refs}
}

\end{document}